# Status and Future Prospects of the Standardization Framework Industry 4.0: A European Perspective

Olga Meyer[1], Marvin Böll[2], and Christoph Legat[2,3] [0000-0002-5843-1845]

[1] Fraunhofer Institute for Manufacturing Engineering and Automation (IPA)
`olga.meyer@ipa.fraunhofer.de`
[2] German Commission for Electrotechnical, Electronic, and Information Technologies, Offenbach. Germany
`marvin.boell@vde.com`
[3] Technical University of Applied Sciences Augsburg, Germany

**Abstract.** The rapid development of Industry 4.0 technologies requires robust and comprehensive standardization to ensure interoperability, safety and efficiency in the Industry of the Future. This paper examines the fundamental role and functionality of standardization, with a particular focus on its importance in Europe's regulatory framework.

Based on this, selected topics in context of standardization activities in context intelligent manufacturing and digital twins are highlighted and, by that, an overview of the Industry 4.0 standards framework is provided. This paper serves both as an informative guide to the existing standards in Industry 4.0 with respect to Artificial Intelligence and Digital Twins, and as a call to action for increased cooperation between standardization bodies and the research community. By fostering such collaboration, we aim to facilitate the continued development and implementation of standards that will drive innovation and progress in the manufacturing sector.

## 1   Introduction

The field of automation engineering has witnessed a remarkable transformation over the past few decades. Rapid advancements in information and communication technology (ICT) have led to an increasing adoption of sophisticated technologies into automation systems. This evolution has not only enhanced the key performance indicators of mechatronic systems like efficiency and productivity of processes but has also opened new possibilities for innovation and growth. However, as automation technologies continue to evolve, so too does the regulatory landscape that governs their deployment and operation.
In today's world, automation is no longer confined to the realms of factories and manufacturing; it has expanded its reach into various sectors, including healthcare, transportation, energy, and infrastructure. This widespread adoption of automation technologies, coupled with the increasing complexity of ICT systems, has brought about a pressing need to address the regulatory challenges that accompany this technological evolution. The convergence of advanced ICT with automation engineering is a double-edged sword. On one hand, it promises unprecedented levels of efficiency, flexibility, and innovation; on the other hand, it introduces a host of new risks and challenges that require careful management. The need for regulations in this context becomes evident as they provide a general framework for ensuring (among others) the safe, reliable, and ethical use of automation systems. Standards play a crucial role in this context defining the state of the technology and, by that, offers a guide to developing legally compliant technical systems and international, global markets [1]. However, an ever-increasing jungle of standards results in high complexity for practitioners, lecturers, and students.

Two technologies are particularly noteworthy because of current developments: Digital Twins and Artificial Intelligence. Both topics are being discussed particularly intensively in the context of Industry 4.0 and are a very active area of research. Furthermore, numerous norms and standards are also being developed in this area on international level but also in national contexts; the regulation of Artificial Intelligence is being discussed worldwide.

In the context of national and international activities, there is a notable absence of consideration of the systemic connection between norms, standards, and regulation. Moreover, the resulting close link between research and standardization is frequently overlooked. Various initiatives of regulatory bodies [2, 3] and standardization bodies [4, 5] as well as research projects [6] are still advertising and practicing the connection of standardization, research and education. Nevertheless, there is even evidence of an increasing separation between the activities of standardization, standardization, and research.

This paper aims to elucidate the interconnections between standardization, regulation, and research from a European perspective. It also provides an overview of current activities in the standardization of artificial intelligence and digital twins in Industry 4.0[1]. Furthermore, it serves as a call to action to strengthen exchange and cooperation.

The remainder of the paper is structured as follows: In the subsequent Section 2, the German and European perspective on standards and norms as well as their differences is discussed. By that, the role of norms and standards within the European legal framework is presented in detail, and finally the structure and interplay of full-consensus standardization on international, European and national level is presented. Afterwards, in Section 3, norms and standards for industrial interoperability by means of Digital Twins and Artificial Intelligence is presented. The section is concluded by a discussion of the convergence of Digital Twins and Artificial Intelligence in the industrial context. Section 4 concludes the article with a summary and outlook.

---

[1] Most of the content is based on the German standardization roadmaps for Industry 4.0 and German standardization roadmap on Artificial Intelligence, in which the authors of this article played a significant role.

## 2 European Perspective: Standards and Norms in a Nutshell

This section commences with an examination of the key terms "norm," "standard," and "consensus level". Afterwards, it is then proceeded to delineate the role of norms and standards within the European legal framework.

### 2.1 Terminology

The principles of standardization in the Federal Republic of Germany are governed by the DIN 820 norm series which delineates a conceptual distinction between technical specifications based on the level of consensus and the involvement of all interested parties. A norm is the outcome of a process in which tangible and intangible objects are standardized through consensus for the public's benefit. This involves the active participation of all interested parties in a planned and collaborative manner, ensuring public involvement to achieve the highest possible acceptance of the results. Therefore, norms are also referred to as full-consensus standards. Conversely, standardization refers to the establishment of technical rules without the mandatory involvement of all interested parties, public participation, or the requirement to reach a consensus.

Nevertheless, this describes the situation in Germany and is not generally valid for other countries or regions. Furthermore, the terminological separation between norms and standards is not spread and consistently used within the international discussions. In contrast, the term standardization is used mostly for any kind of specification [1, 4, 7]. Unfortunately, the underlying consensus approach is often not precisely defined.

### 2.2 Norms and Standards in the European legal framework

Germany and any other country belonging to the European Union run legal systems that follow the basic structure described in this section. In the following, the German and European systems is used as examples without loss of generality. **Fig. 1** provides an overview.

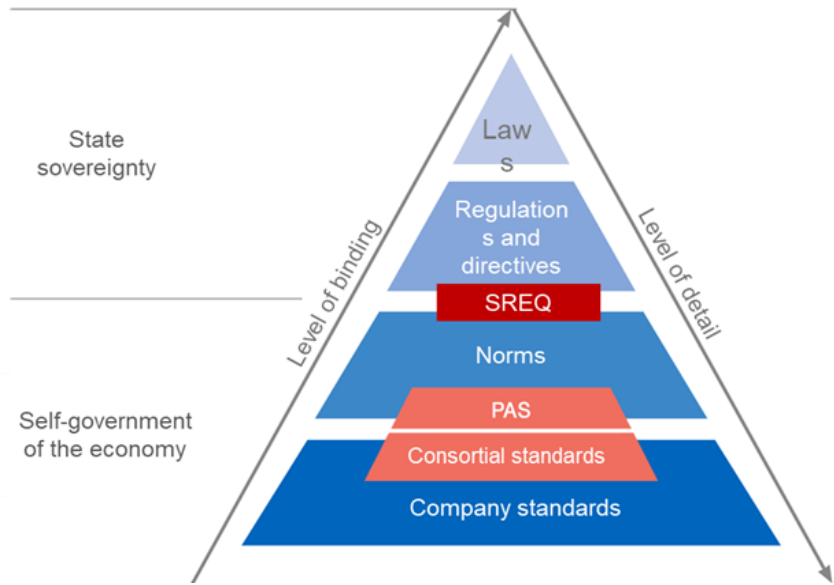

**Fig. 1.** Standards and norms in the legal system of Europe

Norms and standards are closely linked to the legal system. Therefore, it is necessary to take a detailed look at legislation and its mechanisms to gain a fundamental understanding of the role of norms and standards within the legal system.

In Germany, the constitution, known as the Basic Law, is at the top. This is followed by formal laws, which are passed by the parliamentary legislature in accordance with the procedure laid down in the Basic Law. Ordinances, on the other hand, are issued by the executive, i.e. the government, based on an authorization granted by formal law. The content, scope and purpose of the authorization granted must be sufficiently defined in the corresponding formal law, so that no ordinance is possible without a basis of authorization by a higher law. This should be distinguished from general administrative regulations, which are not legal acts addressed to citizens, but administrative acts addressed to the administration and issued by higher bodies within the administrative hierarchy.

The European Union (EU) is a political institution that is central to Europe and functions as an independent legal entity. It is an association of 27 European states and its political system contains both supranational and intergovernmental elements. This is reflected in the organization of the institutions of the European Union. The European Council and the Council of the European Union represent the individual states through their respective governments. The European Parliament is the legislative body of the EU and represents the citizens of the European Union. The European Commission serves as the executive branch, and the Court of Justice of the EU represents the judiciary as a supranational institution.

In European law, there is a distinction between directives and regulations. Regulations are legal acts of the European Union that have general application and direct effect in the member states. Directives are also legal acts of the European Union, but they must be transposed into the national law of the member states by a specified date. It is possible to make minor amendments to them. Directives are typically adopted jointly by the Council of the European Union and the European Parliament through the ordinary legislative procedure, based on a proposal from the European Commission.

In summary, when products are placed on the market or services are provided in Europe, they must comply with both European Union regulations and national laws. These laws can be either of national origin or serve to implement European directives. This is collectively referred to as the legal framework. The legal framework typically outlines specific objectives or necessary general measures but does not dictate the exact methods for achieving these objectives. This is because necessary and appropriate measures, especially in technical contexts, depend heavily on the current state of the art. If these measures were to be included in legislation, regular updates to the legislation would be required to account for changes in the state of the art. Therefore, legislation should remain focused on technology-independent objectives and requirements, while full-consensus standardization should define the current state of the art to be used in this context. Full-consensus standardization is the only entity capable of fulfilling this role, due to the extensive consensus involving all relevant parties on which it is based.

However, this situation is market specific. Other trading markets have different legal structures - but are also subject to some form of regulation. Standards play an important role there as well, albeit in a slightly different way in some cases. In China, for example, there is nothing comparable to the European legal framework - rather, norms and standards are defined as mandatory for the Chinese domestic market. But here, as in Europe, standards are fundamental to making products and services available in a market.

Accordingly, standardization play a dedicated role in context of legislation and regulation because, (in most countries and markets) they are contracted, asked and/or prompted to develop standards defining the state of technology to comply with legal and regulatory frameworks. Even there are slight differences across the world, this basic rule is largely valid in the abstract, even if it is implemented differently in different countries, cultures, and markets.

### 2.3 Full-consensus standardization framework.

National standardization bodies (NSBs) and their respective mirror committees play an important role in representing the country-specific activities and interests of the European standardization organizations (ESOs), namely CEN, CENELEC and ETSI. Similarly, the national committees of European countries are designed to mirror the international committees of ISO, IEC, and ITU. It is worth noting that the communication channels in Europe currently run exclusively via the national mirror committees. This is largely due to the voting rights in the consensus-building of standards at the international level. For this reason, it is common for both an international and (if available) a corresponding European committee to be mirrored by the same national technical committee. This can result in a lack of a consolidated European perspective, coordination, and direct communication from the European level to the international standardization bodies and committees. However, work at the European level, usually in the form of publications, provides valuable guidance for national work. In some cases, there is also an exchange of ideas and opinion-forming at the European level with feedback to the national level.

European standards[2] are developed in several different ways, with CEN (in the electrotechnical field), CENELEC (in the general standards field) and ETSI (in the telecommunications field) all playing a part. CEN and CENELEC have adopted the voting weighting system laid down for the European Union in the Treaty of Nice [8], which is based on the population of the individual countries. All CEN members are obliged to adopt European standards unchanged as national standards and to withdraw all conflicting national standards. In this way, all CEN/CENELEC members apply the same European standards. This is one of the foundations of the European Single Market. The use of European standards is voluntary. As globalization increases, experts are developing standards at the international level. International standards can also be adopted as European standards. In accordance with the Vienna Convention [9], a standard can also be developed either at the international (e.g. ISO) or European (e.g. CEN) level and then adopted simultaneously as an international and European standard through parallel coordination. We are pleased to note that the European standardization organizations are officially recognized as providers of European standards by Regulation (EU) No. 1025/2012 [10]. CEN, CENELEC, and ETSI have had the privilege of working with the European Commission since 1984, when a cooperation agreement was signed. The agreement, revised in 2003, sets out the general guidelines for cooperation.

DIN and DKE, the German NSBs, are among the CEN and CENELEC members with the greatest voting power. DIN, the German Institute for Standardization, plays an important role in standardization in Germany. The DKE, the German Commission for Electrical, Electronic & Information Technologies within DIN and VDE, is the German standards organization responsible for the development and adoption of standards and safety regulations in the fields of electrical, electronic, and information technologies.

CEN, CENELEC and ETSI have been fortunate to have enjoyed a fruitful collaboration with the European Commission since 1984, when a cooperation agreement was signed. This agreement, which was revised in 2003, sets out the general guidelines for cooperation.

## 3    Digital Twins and Artificial Intelligence for Industrial Interoperability

Interoperability is often defined at the communication layers for information and data exchange, but not all models have a dedicated layer for semantic interoperability, which includes both semantic and syntactic aspects. Semantic interoperability ensures that the meaning and formats of exchanged data are preserved between machines and processes, enabling seamless communication across domains. This interoperability extends to human-machine interfaces, where machine interpretations differ from human interpretations. Machines need structured languages and rules to interpret and control processes, while humans can understand mathematically precise descriptions if they are qualified. Designing Industry 4.0 systems to be user-friendly and conducive to learning requires comprehensible information flows and human-system interfaces, supported by appropriate assistance systems that enhance workers' understanding and contextual awareness.

---

[2] European standards are identified by the letters "EN" (European Norm) in their name.

The need to semantically label requested characteristics primarily concerns standardization, requiring the digitalization of standards into machine-interpretable documents. This involves developing suitable semantic labels, or ontologies, and appropriate tools and processes for generating, managing, and applying content. Effective anchoring of these changes in stakeholder infrastructures is crucial, requiring early preparation to adapt and expand accordingly. Standardization aims to provide artefacts for an Industrie 4.0 ontology, integrating both operational and declarative representations. The semiotic domains—semantics/human, ontology/standards, and thing/asset—define the architecture, cooperative behavior, and data types for industrial products, with relationships between these domains forming the semantic assignments necessary for interoperability.

### 3.1 Digital Twins

The Digital Twin is a digital representation of a cyber-physical system (asset) that autonomously mirrors the behavior of its physical counterpart through semantic means. It provides information on various life cycle phases and allows for the analysis, validation, or simulation of the asset's behavior. The results can be fed back to the physical asset for adjustments. Standardization aims to provide the necessary semantic artifacts for building executable models for analysis and simulation. These tools enable the Digital Twin to organize and structure a data space compatible with cyber-physical assets. Current full-consensus standards, like those from ISO/IEC JTC 1/SC 41/WG 6 [11], have yet to fully address the relationship between semantics and the Digital Twin, which is crucial for technical implementation and normative support.

In Industry 4.0 production, interactions between Digital Twins are essential, promoting flexibility, modularization, decentralization, and asset autonomy. The Digital Twin handles the information technology aspects, necessitating standardized exchanges of Industry 4.0 components, facilitated by the I4.0 language defined in VDI/VDE 2193-1 [12]. This language enables the exchange of ontology-required vocabulary and interoperable behavior of components, recommended for international standardization. Semantic standards comprise operational or declarative semantics, I4.0 language for normative requirements, and machine-specific I4.0 engineering elements. The Digital Twin, within the semantic domain, also becomes an administrable asset through the Asset Administration Shell [13–15].

The Asset Administration Shell (AAS) is based on the idea of semantic interoperability. It enables machines, devices, and sensors to interact with each other and communicate with people in Industry 4.0. This results in intelligent Industry 4.0 systems that connect physical objects with data and intelligence to function in a digital ecosystem. Modeling tools are essential for making the properties of the real world available in the information world. They must offer the necessary flexibility for data exchange between assets, especially when heterogeneous manufacturer data is exchanged in the upgrade network. This is often referred to as a universal "integration connector", which is used for data exchange along the value chain to establish interoperability in the digital ecosystems.

In the Asset Administration Shell, the data of the information world is not stored separately. Instead, it is structured in the form of submodels with the properties defined there. This allows us to process, model, retrieve, find, or forward information in a standardized manner in accordance with the interoperability requirements of a digital ecosystem. This means that each Asset Administration Shell is clearly and adequately described in its relationships to other Asset Administration Shells through its submodels. Furthermore, the properties in the Asset Administration Shell can be clearly assigned to an asset. The asset and the associated Asset Administration Shell together form the "Industry 4.0 component" [16].

### 3.2   Artificial Intelligence in Industrial Automation

Germany and Europe are committed to the standardization of AI. This is not least because of its importance in the European legal framework (see Section 2). This is evident in the German government's national AI strategy [17] and the European Commission's strategy and regulatory activities [18]. There is already a wealth of AI standards and specifications in place, developed at the national and international levels, on a consortium basis, and in full consensus. Regulatory and legislative authorities around the world are taking note of the disruptive nature of AI and its impact on the existing legal framework. They are concerned about the correct interpretation and validity of the legal framework in the application of AI.

Since 2022, there has been a growing consensus on a definition of artificial intelligence, which was reached through extensive global discussions. An AI system is a "technical system that produces results such as content, predictions, recommendations or decisions for a given set of human-defined goals" (see ISO/IEC 22989 [19]). However, this definition still has limitations and does not fully address the existing challenge of differing definitions. In Germany, the Joint Committee on Artificial Intelligence of DIN/DKE (NA 043-01-42 GA 65) is responsible for developing national standards in this area. At the European level, the CEN/CENELEC JTC 21 "Artificial Intelligence" was established in June 2021. Its national work in Germany also takes place in NA 043-01-42 GA. This has made it possible to strategically bundle national activities and incorporate them into European and international standardization work. It would be beneficial to intensify efforts to orchestrate, consolidate, and harmonize standardization in the context of AI at the European level, particularly as part of the European regulatory activities that deal with the European Commission's standardization mandate to the European standardization institutes.

### 3.3 Convergence of Digital Twins and Artificial Intelligence

Symbolic Artificial Intelligence is a branch of AI research that emphasizes the use of explicit, human-readable symbols and rules to represent knowledge and reasoning about the world. At the heart of symbolic AI are knowledge graphs, ontologies, and logic-based reasoning. Knowledge graphs are structured representations of facts and relationships between entities that enable machines to understand and navigate complex information networks. Ontologies provide a formalized framework for defining the types, properties, and relationships of entities within a given domain and ensure consistency and interoperability between different systems and applications. Logic-based reasoning, a core component of symbolic AI, involves the application of formal logic to derive conclusions from a set of premises. This approach enables transparent, explainable and interpretable decision-making processes, where each conclusion can be traced back to the underlying rules and data. By combining these elements, symbolic AI enables robust and explainable AI systems capable of performing sophisticated tasks such as natural language understanding, semantic search, and automated reasoning.

In industrial context, ontologies are, among others, used to formulate requirements for products or the manufacture of products that can be checked in their context. This requires search criteria (ontology search points) with which clear identification is possible. For example, IEC 63278-1 [15] states that data models are defined on the basis of ontologies. The semantics of the actual data is typically documented by referencing these ontologies.

Ontologies originate from a variety of application areas and are intended for use in specific contexts. Examples of such ontologies include material science ontologies that facilitate the selection of materials for specific applications based on standard properties, ontologies in the construction industry or product data dictionaries that standardize the representation of products in electronic catalogs. For ontology-based data to be optimally effective in the planning and documentation of production systems, it is essential that the underlying ontologies meet certain minimum quality requirements. For instance, the concepts defined in the ontology must be uniquely identified on a global scale, or special properties must be defined on the basis of clearly specified data types and compatible physical units (see IEC 62832 series [20] and IEC 61360 [21]). The vocabulary defined by those dictionaries represent one bridge between Digital Twins and Artificial Intelligence: Property-based description of the Asset Administration Shell relies on semantic links to dictionary items; dictionaries form ontologies to reason about domain vocabulary which can be used by Artificial Intelligence. Consequently, symbolic AI can reason about (annotated) data provided by Digital Twins. Vice versa, insights derived from symbolic AI can be offered and utilized as a component of an asset administration shell, which may be conceptualized as a Digital Twin.

Semantically annotated data as provided by Asset Administration Shells play also an important role in context of Large Language Models and Generative AI as it is becoming increasingly obvious that these subsymbolic AI methods can be strengthened by combining them with symbolic AI (as provided by Digital Twins / Asset Administration Shells).

## 4      Summary and Outlook

This paper addresses the critical need for robust and comprehensive standardization in the rapidly evolving landscape of Industry 4.0 technologies and emphasizes the importance of interoperability and efficiency for future industry. The discussion centers on the foundational role of standardization in the European regulatory framework and emphasizes its significance in maintaining cohesion and reliability in smart manufacturing and digital twin applications. By examining selected topics within these areas, the paper provides an insightful overview of the current framework for Industry 4.0 standards. It serves as an informative guide detailing the existing standards for artificial intelligence (AI) and digital twins, while advocating for increased collaboration between standardization bodies and the research community. Such collaboration is considered essential to drive the continued development and implementation of standards that support innovation and progress in the manufacturing sector.

In light of the ongoing development of Industry 4.0 technologies, it is becoming increasingly clear that a proactive approach to standardization is required. This paper highlights the necessity for increased collaboration between standards organizations and researchers to address the challenges and opportunities that are emerging. Future efforts should focus on developing adaptable and forward-looking standards that consider the rapid advances in AI, digital twins and other Industry 4.0 technologies. Furthermore, there is a necessity for global harmonization of standards to ensure seamless interoperability across international borders and to facilitate the development of a truly global Industry 4.0 digital ecosystem. By strengthening synergies between regulatory frameworks, industry stakeholders, and the research community, we can pave the way for a more innovative, efficient, and secure industrial future. This collaborative approach will be critical to realizing the full potential of Industry 4.0 technologies and ultimately achieving significant advances in manufacturing and beyond.

**Acknowledgements**

This work was supported by the European Union's Horizon Europe research and innovation program under grant agreement No. 101070229 – STAND4EU.